\documentclass[%
reprint,
amsmath,amssymb,
prb,
]{revtex4-1}

\usepackage{graphicx}
\usepackage{dcolumn}
\usepackage{bm}
\usepackage{multirow} 

\usepackage{siunitx}

\newcommand{\RN}[1]{\uppercase\expandafter{\romannumeral#1}}

\begin{document}

\title{Surface acoustic wave coupling between micromechanical resonators}

\author{Hendrik K\"ahler}
\author{Daniel Platz}
\author{Silvan Schmid}%
 \email{silvan.schmid@tuwien.ac.at}
\affiliation{%
 Institute of Sensor and Actuator Systems, TU Wien, Gusshausstrasse 27-29, 1040 Vienna, Austria.
}%

\date{\today}

\begin{abstract}
The coupling of micro- or nanomechanical resonators via a shared substrate is intensively exploited to built systems for fundamental studies and practical applications. So far, the focus has been on devices operating in the kHz regime with a spring-like coupling. At resonance frequencies above several 10 MHz, wave propagation in the solid substrate becomes relevant. The resonators act as sources for surface acoustic waves (SAWs), and it is unknown how this affects the coupling between them. Here, we present a model for MHz frequency resonators interacting by SAWs and derive the eigenfrequencies and quality factors of a pair of resonators for the symmetric and antisymmetric mode. Our results are in agreement with finite element method (FEM) simulations and show that, in contrast to the well-known strain-induced spring-like coupling, the coupling via SAWs is not only dispersive but also dissipative. This can be exploited to realize high quality phonon cavities, an alternative to acoustic radiation shielding by, e.g. phononic crystals. 
\end{abstract}

\maketitle

\begin{figure}[t]
  \begin{center}    
    \includegraphics{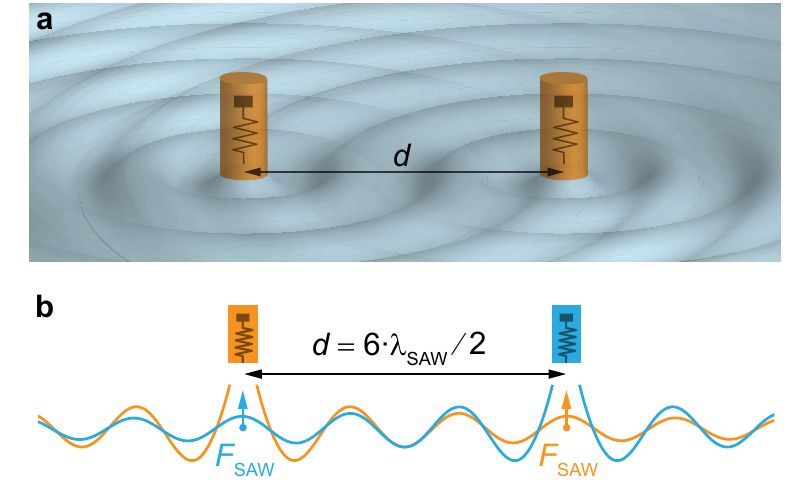}
    \caption{Pillar resonators coupled by SAWs. \textbf{a} Schematic of the symmetric mode of two SAW-coupled identical pillar resonators. The pillar resonators vibrate in out of plane direction (first bulk mode) and act as sources for SAWs. \textbf{b} Schematic of the SAW coupling mechanism illustrating the effective forces $F_\text{SAW}$ exerted on the pillars resonators by the emitted SAWs. The pillar resonators vibrate in the symmetric mode and are shown in the moment of maximum displacement.}
    \label{WCTheory}
  \end{center}    
\end{figure}

Two mechanical resonators mounted on the same substrate are considered coupled if the motion of one of the resonators affects the motion of the other resonator and vice versa. Such systems of coupled micro- or nanomechanical resonators are widely used in fundamental studies and practical applications. They are utilized to build highly precise mass sensors \cite{Spletzer2006,Gil-Santos2009,Stassi2019}, allow for the study of quantum-coherent coupling and entanglement between two distinct macroscopic mechanical objects \cite{Kotler2021,Okamoto2013}, and enable the investigation of collective dynamics \cite{Doster2019}. In all of these cases, the considered resonance frequencies are in the kHz or low MHz regime. Here, we focus on micro- and nanomechanical resonators with resonance frequencies in the order of several \SI{10}{\mega\hertz} and higher. At these frequencies, the wavelength of surface acoustic waves (SAWs) becomes much smaller than the usual dimensions of the resonators' substrates, assuming a typical SAW velocity  \cite{Morgan2007,Fu2017} of around \SI{3000}{\meter / \second}. This allows a coupling of the resonators by SAWs. An example for SAW-coupled resonators is illustrated in Fig.~\ref{WCTheory}a by a pair of short micro-pillar resonators. Typical resonance frequencies of such pillar resonators are above \SI{50}{\mega\hertz}. Short micro-pillar resonators are compatible with SAW devices and are utilized to manipulate the propagation of SAWs \cite{Benchabane2017,Bonhomme2019,Achaoui2011,Liu2019,Colombi2016}.

In a lumped-element model, a substrate-mediated coupling is usually represented by a spring connecting two spring-mass systems \cite{Gil-Santos2011,Schmid2016}. In doing so, the interaction between the resonators is assumed to be instantaneous. This assumption is not valid if the resonators interact by SAWs. In this case, the coupling between the resonators has a delay. This is the propagation time of a SAW, which is created by one resonator \cite{Berte2018,Jin2017} and travels to the other. Delay-coupled systems have been studied before: two resonators mounted on a string \cite{Lepri2014}, two resonators coupled by a rod \cite{Edelman2013}, or the interaction of air bubbles in water \cite{Feuillade1995,Ooi2008,Feuillade2001,Doinikov2005}, for example. However, a radiative coupling by SAWs between three dimensional micro- or nanomechanical resonators mounted on a semi-infinite substrate has not yet been investigated. A coupling function, analogous to the spring coupling model, is unknown and would be helpful to deeper understand recent experiments of coupled micro-pillar resonators \cite{Raguin2019}.

Here, we propose a coupling function for SAW-coupled micro- or nanomechanical resonators. We analytically derive the eigenfrequencies and quality factors of the symmetric and antisymmetric mode of a pair of coupled resonators as a function of their distance and compare the results to FEM simulations. For the FEM simulations we chose a pair of pillar resonators vibrating in the first bulk mode, since it represents the most simple case. However, the proposed model applies also to other vibrational modes and resonator geometries. As two examples, we consider the second vibrational mode of the pillar resonators and a typical thin cantilever geometry in Supplementary Notes~3 and 4. Finally, we discuss the strength of the SAW-coupling between the pillar resonators. \\
\\
\large
\textbf{Results} 
\newline
\normalsize
\textbf{SAW coupling model.}
The coupling by SAWs is schematically depicted in Fig.~\ref{WCTheory}b. Each of the resonators emits a SAW, which exerts an effective force $F_\text{SAW}$ on the other resonator. The origin of $F_\text{SAW}$ is the mechanical vibration of the resonators and the resulting forces exerted on the substrate surface. In a lumped-element model, the effective force exerted on the substrate by a single resonator with an effective mass $m$ and a displacement $z$ is given by
\begin{equation}
F_\text{s}(t) = - g \, m \, \ddot{z} \; ,
\label{Fs}
\end{equation}
based on the principle of action and reaction. The dimensionless parameter $g$ represents the acoustic coupling strength between the resonator and the substrate. It is not the force $F_\text{s}$ that directly acts on another resonator but the SAW created by $F_\text{s}$. Due to the propagation of the SAW, the force $F_\text{SAW}$ exerted by the SAW has a smaller amplitude and is phase-shifted compared to $F_\text{s}$.
The change in the amplitude of $F_\text{SAW}$  corresponds to the change of the SAW amplitude and the phase shift is the difference in phase between the resonator and it's emitted SAW at the site of the other resonator. Consequently, the force exerted on the other resonator by the emitted SAW is given by
\begin{equation}
F_\text{SAW}(d,t) =  A_\text{SAW}(d) \, F_\text{s}(t) \, e^{-i \, \Delta\phi(d,\omega)} 
\label{FSAW}
\end{equation}
with
\begin{equation}
\begin{split}
\Delta\phi(d,\omega) &= \phi_\text{R} - \phi_\text{SAW}(d,\omega) \\
&= \frac{2\pi \, \omega}{c_\text{SAW}} \, d + \theta(d) \; , 
\end{split}
\label{Phase}
\end{equation}
where $A_\text{SAW}(d)$ and $\phi_\text{SAW}$ are the normalized ($A_\text{SAW}(0) = 1$) amplitude and phase of the emitted SAW at the site of the other resonator, $\phi_\text{R}$ and $\omega$ are the phase and vibration frequency of the SAW-emitting resonator, $c_\text{SAW}$ is the velocity of the emitted SAW, and $\theta$ represents additional phase changes of the emitted SAW during its formation (see Supplementary Note~1 for details). 

In Eq.~(\ref{FSAW}), we do not give an explicit expression for the SAW amplitude, since the amplitude of a SAW emitted by a three dimensional resonator mounted on a semi-infinite substrate strongly depends on the materials. First, the materials affect the distribution of the radiated energy into bulk acoustic waves (BAWs) and SAWs \cite{Shibayama1976}. Second, some substrate materials are anisotropic, such as single crystalline silicon or lithium niobate, so that the intensity of the SAWs depends on the propagation direction.\\
\\
\textbf{Two SAW-coupled resonators.}
In the following, we consider the symmetric and antisymmetric mode of two identical resonators coupled by SAWs and derive the modes' eigenfrequencies and quality factors as a function of the resonators' distance $d$. Two SAW-coupled resonators exert the force $F_\text{SAW}$ on each other. For weak damping, this results in the following equations of motion
\begin{align}
\ddot{z_1}+ \frac{\omega_0}{Q_0} \, \dot{z_1}+\omega_0^2 \,z_1 + g \, A_\text{SAW}(d) \, \ddot{z_2} \, e^{-i \, \Delta\phi(d,\omega_2)} &= 0 \, , \label{Derive1a}  \\
\ddot{z_2}+ \frac{\omega_0}{Q_0} \, \dot{z_2}+\omega_0^2 \,z_2 + g \, A_\text{SAW}(d) \, \ddot{z_1} \, e^{-i \, \Delta\phi(d,\omega_1)} &= 0 \, ,
\label{Derive1b}
\end{align}
where the indices 1 \& 2 give the number of the resonator and  $\omega_0 = 2\pi \, f_0$ and $Q_0$ are the eigenfrequency and quality factor of a single resonator, respectively. If the two resonators vibrate in a symmetric ($+$) or antisymmetric ($-$) mode, we can drop the indices, since the modes feature $\ddot{z}_\text{1} = \pm\ddot{z}_\text{2}$. The equations of motion of both resonators are then given by
\begin{equation}
\ddot{z}+ \frac{\omega_0}{Q_0} \, \dot{z}+\omega_0^2 \,z \pm g \, A_\text{SAW}(d) \, \ddot{z} \, e^{-i \, \Delta\phi(d,\omega)} = 0 \; .
\label{Derive2}
\end{equation}
Using the ansatz $z = z_0 \, e^{i \, \omega \,t}$ with complex amplitude $z_0$ results in
\begin{equation}
 [-\omega^2 + i \, \omega \, \frac{\omega_\pm(d)}{Q_\pm(d)} + \omega_\pm^2(d)] \, z = 0
\label{Derive3}
\end{equation}
with
\begin{equation}
 \omega_\pm(d) = \frac{\omega_\text{0}}{\sqrt{1 \pm g \, A_\text{SAW}(d) \, \text{cos} \big( \Delta\phi(d,\omega_\pm) \big)}} 
\label{fsym}
\end{equation}
and
\begin{equation}
\begin{split}
 Q_\pm(d) = \frac{\frac{\omega_0}{\omega_\pm(d)} \, Q_0}{1 \pm \frac{\omega_\pm(d)}{\omega_0} \, Q_0 \, g \, A_\text{SAW}(d) \, \text{sin} \big( \Delta\phi(d,\omega_\pm) \big)} \;.
\end{split} 
\label{Qsym}
\end{equation}
By comparing Eq.~(\ref{Derive3}) with the case of a single resonator, it becomes clear that $\omega_\pm$ and $Q_\pm$ are the eigenfrequencies and quality factors of the symmetric ($+$) and antisymmetric ($-$) modes of the SAW-coupled resonators. In contrast to the spring-like coupling, the coupling by SAWs not only modulates the resonators' eigenfrequencies (dispersive coupling) but also modulates their damping (dissipative coupling). Due to the dependency of $\Delta \phi$ on the resonators' vibration frequency, $\omega_\pm$ is a function of itself. In case of small modulations of the eigenfrequencies $\omega_\pm \approx \omega_0$, the eigenfrequencies $\omega_\pm$ can be approximated by using $\Delta\phi(d,\omega_\pm) \approx \Delta\phi(d,\omega_0)$. Furthermore, it is important to note that the SAWs are a part of the resonator's radiation losses into the substrate, which are represented by $Q_\text{rad}$. Hence, the coupling force $F_\text{SAW}$ can only modulate $Q_\text{rad}$ and not other damping mechanism included in $Q_\text{0}$. Consequently, the product $Q_\text{0} \, g$ in Eq.~(\ref{Qsym}) must be proportional to $Q_\text{0}/Q_\text{rad}$, which gives $g \propto 1/Q_\text{rad}$. This matches with the definition of g as the acoustic coupling strength between the resonators and the substrate.\\
\\
\textbf{FEM model.} To test the proposed SAW coupling model, we performed FEM simulations of a pair of coupled pillar resonators, which are mounted on a semi-infinite substrate and vibrate in the first bulk mode. We simulated the eigenfrequencies and quality factors of the symmetric and antisymmetric modes of the pillars as a function of the spacing between them. The pillars have a diameter of \SI{4}{\micro\meter} and a height of \SI{6}{\micro\meter}. Their Young's modulus, mass density, and Poisson's ratio are  $E = 4.88~\text{GPa}$, $\rho = 1183~\text{kg/m}^3$, and $\nu = 0.22$, similar to the material properties of SU-8 resist. No intrinsic damping was considered, since the focus was on the modulation of the radiation losses into the substrate. The substrate was modeled as a half sphere and partitioned into an outer and an inner part. The outer part was defined as a perfectly matched layer (PML) mimicking an infinitive large substrate. The substrate's  material was lithium niobate (LiNbO$_3$) with a 127.86$^\circ$ Y-cut orientation. The pillars were placed along a line perpendicular to the crystallographic X-axis (geometric y-direction).  Along the line, Rayleigh-type SAWs emitted by a point source propagate with a velocity \cite{Kovacs1990,Holm1996,Laude2008} of around $c_\text{Ray}=\SI{3664}{\meter / \second}$. Simulations of a single pillar show that the pillars emit Rayleigh waves (see Supplementary Note~2).
\begin{figure}[t]
  \begin{center}    
    \includegraphics{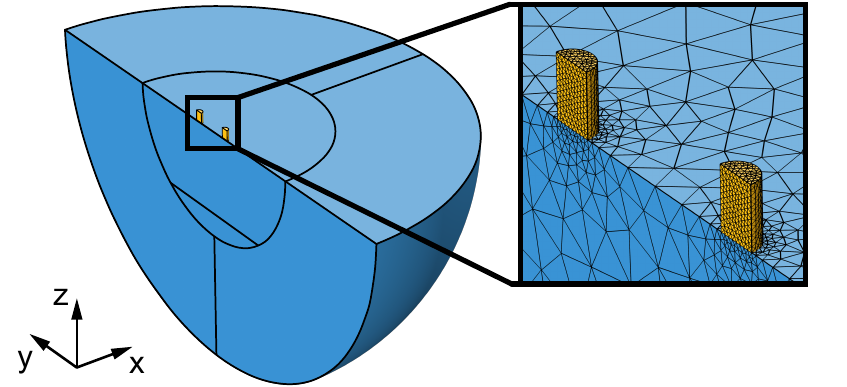} 
    \caption{Geometry and mesh of the FEM simulations of two pillar resonators for a separation distance of $d = \lambda_\text{SAW}/2$ with $\lambda_\text{SAW} = c_\text{Ray}/f_0$.}
    \label{FEMGeometry}
  \end{center}    
\end{figure}
The symmetry of the LiNbO$_3$ crystal\cite{Weis1985} allowed for reducing the simulated domain to half of the considered domain. The geometry and the used mesh are shown in Fig.~\ref{FEMGeometry}. Further details to the FEM simulations are given in the Methods section.\\
\\
\textbf{SAW model vs. FEM simulations.} To compare the proposed SAW coupling model with the FEM simulations, we determined $g$ by fitting Eq.~(\ref{fsym}) to the simulated eigenfrequencies first, then plotted Eq.~(\ref{fsym}) and Eq.~(\ref{Qsym}) together with the simulated eigenfrequencies and quality factors. For the fitting and plotting, we exploited the small modulation of the eigenfrequencies, as discussed above. In addition to the the normalized SAW amplitude $A_\text{SAW}$, we also determined the phase difference $\Delta \phi$ by a FEM simulation of a single pillar resonator. Along a line in y-direction on the surface, we determined the displacement in z-direction $u_\text{z}$ of an SAW emitted by the single pillar and calculated the absolute value $\left| u_\text{z} \right| \propto A_\text{SAW}$ and the phase ${\phi_u}_\text{z} = \phi_\text{SAW}$. We chose the displacement in z-direction, since the pillars vibrate in z-direction. Based on $\phi_\text{SAW}$ and Eq.~(\ref{Phase}), we calculated then $\Delta \phi$, with $\phi_\text{R}$ determined at the central point on top of the single pillar. The results of the single pillar FEM simulation are given in Supplementary Note~2.
\begin{figure}[t]
  \begin{center}    
    \includegraphics{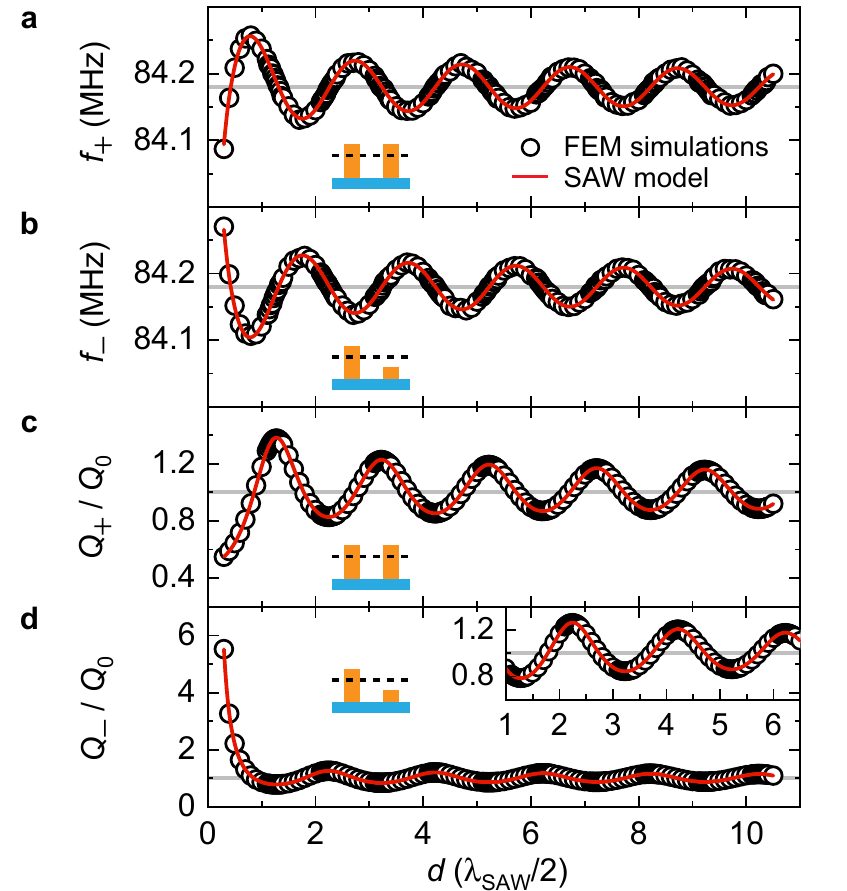}
    \caption{FEM simulations of a pair of pillar resonators. Eigenfrequency and quality factor of the symmetric and antisymmetric modes as a function of the distance $d$ between the pillars. The grey lines mark the properties $f_0 = 84.18~\text{MHz}$ and $Q_0 = 208$ of a single pillar resonator. The red lines are plots of Eq.~(\ref{fsym}) and Eq.~(\ref{Qsym}) for $g=0.027$ and exploiting $\omega_\pm \approx \omega_0$.}
    \label{FEM2Pillars}
  \end{center}    
\end{figure}
\begin{figure*}[t]
  \begin{center}    
    \includegraphics{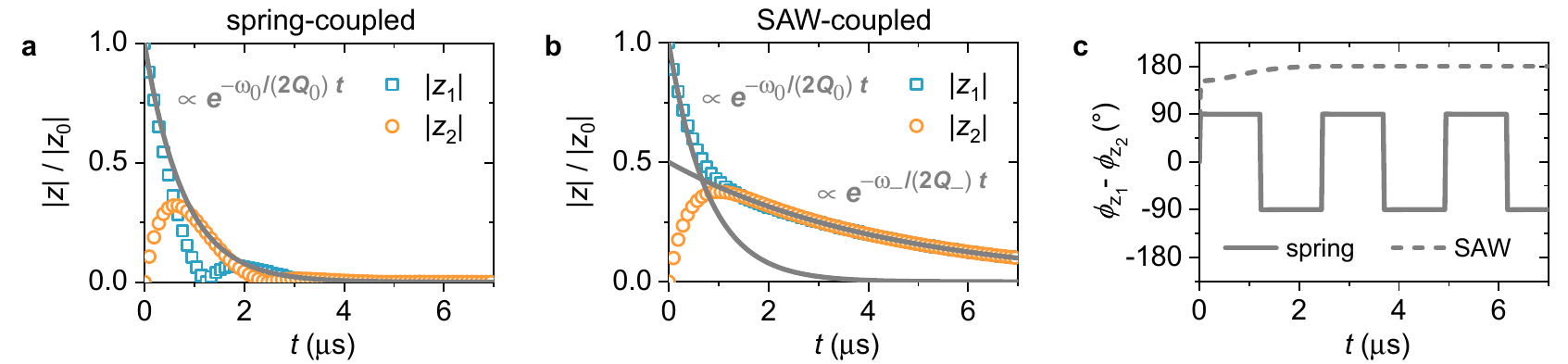} 
    \caption{Dynamics of two spring-coupled resonators at the threshold of the strong coupling regime in comparison to two SAW-coupled pillar resonators separated by $d =0.3 \,\lambda_\text{SAW}/2$. Amplitudes of the two spring-coupled resonators \textbf{a} and the two SAW-coupled pillar resonators \textbf{b} as a function of time $t$. A single spring-coupled resonator has the same eigenfrequency $f_0 = 84.18~\text{MHz}$ and quality factor $Q_0 = 208$ as a single pillar resonator. The grey lines show the amplitude decay of a single pillar resonator and of two antisymmetrically vibrating pillar resonators separated by $d =0.3 \,\lambda_\text{SAW}/2$. In the latter case, the FEM simulations discussed above gave $f_- = 84.27~\text{MHz}$ and $Q_- = 1145$. In both graphs, we reduced the amount of data points for the sake of clarity. \textbf{c} Phase difference between the displacements of the two spring-coupled and the two SAW-coupled pillar resonators.}
    \label{Fig4}
  \end{center}    
\end{figure*}

The simulated eigenfrequencies $f_\pm$ and quality factors $Q_\pm$ of the symmetric and antisymmetric mode of the two pillar resonators are plotted in Fig.~\ref{FEM2Pillars} as a function of the pillars' separation distance. The distances between the pillars range from 0.3 to $10.5 \,\lambda_\text{SAW}/2$. In comparison, coupled beams or cantilevers \cite{Gil-Santos2011,Okamoto2013,Stassi2019} in the frequency regime around \SI{100}{\kilo\hertz} usually have a separation distance in the order of $\SI{e-3} \,\lambda_\text{SAW}/2$, assuming a typical SAW velocity  \cite{Morgan2007,Fu2017} of around \SI{3000}{\meter / \second}. The eigenfrequencies $f_\pm$ and quality factors  $Q_\pm$ of both modes are periodically modulated with increasing distance $d$. The modulations of $f_\pm$ and $Q_\pm$ are phase shifted by $\pi/2$, which implies that the coupling between the pillars is purely dispersive at the maxima and minima of $f_\pm$ and purely dissipative at its zeros. The results of the FEM simulations are in excellent agreement with the proposed SAW coupling model, as can be seen in Fig.~\ref{FEM2Pillars} by the red lines.

With increasing distance $d$, the amplitude of the modulations of $f_\pm$ and $Q_\pm$ decreases, since the amplitude of the emitted SAWs decreases with increasing propagation distance. The huge difference in the modulation strength between $Q_+$ and $Q_-$ at small distances $d$ becomes clear by considering the limiting case $d \rightarrow 0$. The quality factor is defined as the ratio between the energy stored and energy lost during one cycle at resonance \cite{Schmid2016}. In the antisymmetric case, the pillars behave as a nonradiating source \cite{Berry1998,Gbur1999,Miroshnichenko2015}. No power is emitted due to destructive interference of the emitted waves, which results in $Q_- \rightarrow \infty$. In the symmetric case, the two pillars emit four times the power emitted by a single pillar due to constructive interference \cite{Miles2020} and store together twice the total energy, which yields  $Q_+/Q_0 = 0.5$. The case $d \rightarrow 0$ shows that the dissipative coupling can also be interpreted as a result of wave interference, which explains the difference in the modulation strength of the quality factors and the eigenfrequencies. Only a small fraction of the waves emitted by one resonator reaches the other resonator directly. In comparison, the interference affects all waves emitted by the pillars.\\
\\
\textbf{Strength of the SAW coupling.}
A feature of major interest in the context of coupled micro- and nanomechanical resonators is the strength of the coupling. A distinction is typically made between the weak and the strong coupling regime. In the weak coupling regime the energy stored in one resonator dissipates before it is transferred to the other resonator, whereas in the strong coupling regime the resonators exchange energy and the dynamics of the coupled resonators is governed by the motion of both resonators \cite{Rodriguez2016}. For spring-like coupled resonators the strength of the coupling can be estimated by the ratio $(f_- - f_+)/(f_0/Q_0)$, since $f_- - f_+$ is proportional to the energy exchange rate between the resonators and $f_0/Q_0$ is proportional to the energy dissipation rate of the system \cite{Rodriguez2016}. This approach can not be used for SAW-coupled resonators, since the quality factor of the resonators is a function of the time $t$ due to the dissipative part of the coupling. To get an impression of the strength of the coupling between the two pillar resonators, we considered the dynamics of two pillar resonators separated by $d =0.3 \,\lambda_\text{SAW}/2$ in comparison with two spring-coupled resonators at the threshold of the strong coupling regime. In detail, we calculated the displacements of the resonators as a function of time by numerically solving the equations of motion of each system for the following complex initial conditions
\begin{equation}
\begin{aligned}
z_1(t=0) &= |z_0| \; , & \quad z_2(t=0) &= 0 \; ,\\    
\dot{z_1}(t=0) &= i \, \omega_0 \, |z_0| \; , & \quad  \dot{z_2}(t=0) &= 0 \; ,
\label{Initial}
\end{aligned}
\end{equation}
 where $z_0$ is a complex amplitude. At $t=0$, both resonators are at rest, but resonator 1 is displaced. We chose $\dot{z_1}(t=0)$ such that the displacement and the velocity of resonator 1 are phase shifted by $\pi/2$, and that its maximal kinetic energy equals its maximal potential energy.
 
 The equations of motion of two SAW-coupled resonators are given by Eq.~(\ref{Derive1a}) and (\ref{Derive1b}). For slight damping, the equations of motion of two identical spring-coupled resonators are given by
\begin{align}
\ddot{z_1}+ \frac{\omega_0}{Q_0} \, \dot{z_1}+\omega_0^2 \,z_1 + \omega_c^2 \, (z_1-z_2) &= 0 \; ,
\label{SpringCoupa}  \\
\ddot{z_2}+ \frac{\omega_0}{Q_0} \, \dot{z_2}+\omega_0^2 \,z_2 - \omega_c^2 \, (z_1-z_2) &= 0 \; ,
\label{SpringCoupb}
\end{align}
where $\omega_c$ is the coupling rate. We numerically solved both systems for the following parameter values
\begin{align*}
\omega_0& = 84.18~\text{MHz} \; , & A_\text{SAW} &= 0.16 \; ,  & g &=0.027 \; ,  \\
Q_0 &= 208 \; , & \Delta\phi  &= 1.09 \; ,  & \omega_c &= 5.85~\text{MHz} \; .
\label{ParaValues}  
\end{align*}
The given values of $\omega_0$, $Q_0$, $A_\text{SAW}(d)$ and $\Delta\phi(d)$ originate from the FEM simulations discussed above, whereby $A_\text{SAW}(d)$ and $\Delta\phi(d)$ were evaluated at $d =0.3 \,\lambda_\text{SAW}/2$. The value of g was determined by fitting of Eq.~(\ref{fsym}) to the simulated eigenfrequncies of two pillar resonators, as discussed above, and the value of the coupling rate $\omega_c$ results in $(f_- - f_+)/(f_0/Q_0) = 1$, which is the definition of the threshold of the strong coupling regime for spring-coupled resonators \cite{Schmid2016,Doster2019}. Further details to the numerical calculations are given in the Methods section.

The calculated amplitudes of the resonators and their phase relation are depicted in Fig.~\ref{Fig4}. It can be seen that the two systems exhibit different dynamics. The spring-coupled resonators exchange their total energy $E_\text{tot} \propto |z_i|^2$ back and forth and vibrate with a phase difference of $90^\circ$, as shown in Fig.~\ref{Fig4}a,c. In contrast, the pillar resonators vibrate antisymmetrically after a transient time and only transfer energy until both pillars have the same total energy, as shown in Fig.~\ref{Fig4}b,c. The reason for the different dynamics is the dissipative part of the SAW coupling. In contrast to the spring-coupled resonators, the pillar resonators maximize their quality factor, which is displayed by the exponentially decaying lines in Fig.~\ref{Fig4}a,b. Based on the FEM simulations discussed above, two pillar resonators, which are separated by $d =0.3 \,\lambda_\text{SAW}/2$ and vibrate in the antisymmetric mode, have a quality factor of $Q_- = 1455$ instead of $Q_0 = 208$. By comparing the amplitudes $|z_2|$ of both systems, it can be seen that the second pillar resonator has a larger maximum amplitude than the second spring-coupled resonator. Consequently, the ratio of the energy exchange rate to the energy dissipation rate is larger for the SAW-coupled pillar resonators than for the spring-coupled resonators. Since the spring-coupled resonators are at the threshold of the strong coupling regime, the pillar resonators are strongly coupled. It is important to note that the pillar resonators' damping is dominated by radiation losses into the substrate. If this were not the case, the coupling would be weaker.\\
\\
\large
\textbf{Discussion} 
\newline
\normalsize
In conclusion, our results indicate that the coupling between MHz frequency resonators can significantly differ from the usually assumed and purely dispersive spring-like coupling. The coupling via SAWs is not only dispersive, but also dissipative. This makes our results important for all fields working with micro- or nanomechanical resonators with resonance frequencies in the MHz regime and above. For instance, in MEMS and NEMS, a major approach to increase the devices' sensitivity is to shrink the sizes of the mechanical resonators, which shifts their resonance frequencies from the kHz to the MHz regime \cite{Yang2006,Li2007,Dominguez-Medina2018}. Furthermore, for quantum applications at room temperature a high $Qf$-product is needed \cite{Tsaturyan2017,Norte2016}. The dissipative coupling opens up a new possibility to reduce acoustic radiation losses via the substrate by positioning two or more resonators in certain distances. In comparison to acoustic radiation shielding with phononic crystals \cite{Chan2011,Arrangoiz-Arriola2019,Riedinger2016}, phonon cavities based on dissipative coupling are accessible by external SAWs and, thereby, allow to perform quantum acoustics experiments between a SAW \cite{Gustafsson2014} and coupled mechanical resonators.\\ 
\\
\large
\textbf{Methods} 
\newline
\normalsize
\textbf{Details to the FEM simulations.} The FEM simulations were carried out with \mbox{COMSOL} Multiphysics (Version 5.4) using the Modules AC/DC (Electrostatics) and Structural Mechanics (Solid Mechancis, Piezoelectricity). An unstructured tetrahedral mesh was used for the pillars with a maximum element size of an eighth of the pillars' diameter. A swept mesh was utilized for the PML and an unstructured tetrahedral mesh for the inner part of the substrate. The latter had a maximum element size of an eighth of the SAW wavelength $\lambda_\text{SAW}$ up to a distance of $\lambda_\text{SAW}$ to the surface, which is approximately the penetration depth of a SAW \cite{Morgan2007}. For elements deeper in the substrate, the maximum element size was reduced by a factor of two. In Solid Mechanics, quadratic serendipity elements were used as Lagrange elements were used in Electrostatics. We performed an eigenfrequency study to calculate the eigenfrequencies and quality factors of the symmetric and antisymmetric normal modes for different distances $d$ between the pillars and in the simulation of a single pillar resonator.

To minimize the memory requirements and the solution time, we reduced the simulated domain to half of the considered domain by using the following symmetric boundary condition
\begin{equation*}
\mathbf{u} \cdot \mathbf{n} = 0 \; ,
\label{BCsym}
\end{equation*}
where $\mathbf{u}$ is the displacement vector and $\mathbf{n}$ is the unit normal vector of the considered sectional plane. In our case, this is the yz-plane, as shown in Fig.~2. The yz-plane lies in a plane of mirror symmetry of the lithium niobate substrate \cite{Weis1985}, and a pillar resonator made out of an isotropic material shows no displacement in x-direction at any point in the yz-plane when vibrating in the first bulk mode in the limit of small vibrational amplitudes \cite{Schmid2016,Weaver1990}.\\
\\
\textbf{Details to the numerical calculations.} The numerical calculations were carried out in Matlab 2020a. We converted the second-order differential equations to first-order differential equations and solved the first-order differential equations by the function \textit{ode45} using a stepzise of $\Delta t = 25/f_0$, and a relative and absolute error tolerance of $10^{-6}$ and $10^{-8}$, respectively.\\
\\
\large
\textbf{Data availability} 
\newline
\normalsize
The data that support the findings of this study are available from the corresponding author upon reasonable request.\\
\\




\bibliography{PaperBib_Version3}
\vspace{0.1cm}
\large
\noindent \textbf{Acknowledgements} 
\newline
\normalsize
We thank A.L. Gesing for his support with the FEM simulations as well as M. Piller and R. West for many fruitful discussions. This work is supported by the European Research Council under the European Unions Horizon 2020 research and innovation program (Grant Agreement-716087-PLASMECS).\\
\\
\large
\textbf{Authors contribution} 
\newline
\normalsize
H.K. developed the wave coupling model with the support from D.P. and performed the FEM simulations and the numerical calculations including the data analysis. H.K. wrote the paper with input from all authors. The project was supervised by S.S..\\
\\
\large
\textbf{Conflicts of interest} 
\newline
\normalsize
The authors declare no competing interests.

\clearpage

\setcounter{figure}{0} 
\renewcommand\thefigure{S\arabic{figure}}

\setcounter{equation}{0} 
\renewcommand\theequation{S\arabic{equation}}

\onecolumngrid

\section*{Supplementary Information for:\\
Surface acoustic wave coupling between micromechanical resonators}

\vspace{0.6cm}

\twocolumngrid

\subsection*{1. Origin of the phase term $\theta$}

\begin{figure}[t]
  \begin{center}    
    \includegraphics{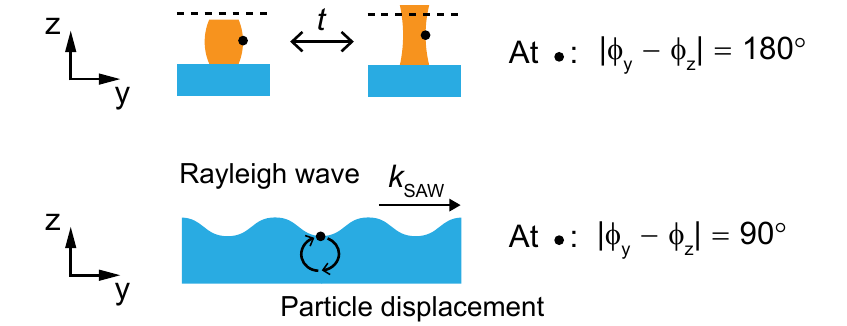} 
    \caption{Schematic of a vibrating pillar resonator and a propagating Rayleigh SAW with the corresponding phase differences between the displacements along the y- and z-direction at the marked spots.}
    \label{OriginTheta}
  \end{center}    
\end{figure}

An explanation for the phase term $\theta$ is shown in Fig.~\ref{OriginTheta}. We consider a system in which a single pillar resonator vibrates in the first bulk mode and emits a SAW. Since the pillar vibrates in the out of plane direction, we assume that the emitted SAW is a Rayleigh wave. 
Inside the vibrating pillar resonator the phase difference between the displacements in y- and z-direction is ideally $180^\circ$ due to the lateral strain. In a Rayleigh wave \cite{Morgan2007} the phase difference between its transversal and longitudinal displacements, here in y- and z-direction, is $90^\circ$. Thus, in order to fulfill the phase relation between the transversal and longitudinal displacements in a Rayleigh wave, the SAW, which has its origin in the pillar, must undergo a phase change in at least one of its displacement directions in addition to the phase change due to propagation.

\subsection*{2. Single pillar FEM simulation (bulk mode)}
\begin{figure}[t]
  \begin{center}    
    \includegraphics{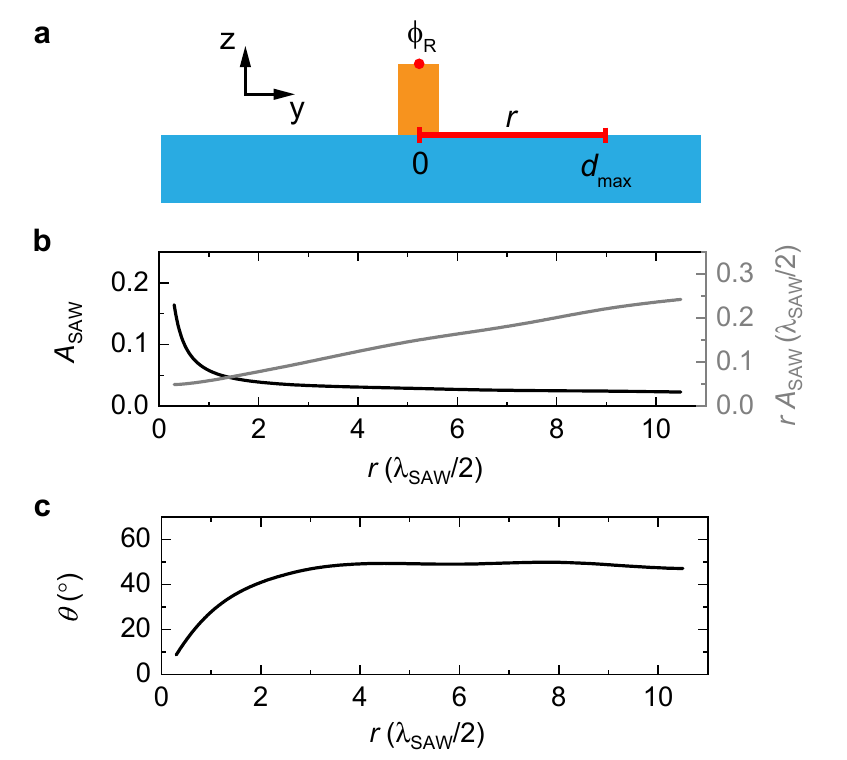} 
    \caption{FEM simulation of a single pillar resonator. \textbf{a} Schematic of a single pillar resonator mounted on a substrate illustrating along which line the normalized amplitude $A_\text{SAW}$ and the additional phase difference $\theta$ of the pillar's emitted SAW were determined. The parameter $d_\text{max}$ is the maximum distance between two pillars in the FEM simulations of a pair of pillar resonators. The red dot marks the spot at which the phase of the resonator $\phi_\text{R}$ was determined. \textbf{b} Normalized amplitude $A_\text{SAW}$ \big($A_\text{SAW}(0) = 1$\big) of a SAW emitted by a single pillar resonator after propagating a distance $r$ along the y-direction. \textbf{c} Additional phase difference $\theta$ between the phase of the pillar resonator and the phase of the pillar's emitted SAW at distances $r$ to the pillar.}
    \label{FEMSinglePillar}
  \end{center}    
\end{figure}

A schematic of the single pillar FEM simulation is shown in  Fig.~\ref{FEMSinglePillar}a. It illustrates the line in y-direction and the reference point for the phase of the resonator, which are mentioned in the main text. As discussed in the main text, we calculated the normalized amplitude $A_\text{SAW}$ and the phase difference between a pillar and its emitted SAW $\Delta \phi$ along the line in y-direction based on the displacement in z-direction $u_\text{z}$ ($\left| u_\text{z} \right| \propto A_\text{SAW}$ and ${\phi_u}_\text{z} = \phi_\text{SAW}$). The determined normalized amplitude $A_\text{SAW}$ is shown in Fig.~\ref{FEMSinglePillar}b. As expected, $A_\text{SAW}$ is not propotional to 1/$r$, since the SAW is bound to the surface and the lithium niobate substrate is not isotropic.

Based on the determined $\phi_\text{SAW}$ we calculated first the phase difference $\Delta \phi$ and then $\theta$, both by using Eq.~(3) in the main text. For the latter, we assumed that the emitted SAW by the pillar resonator is a Rayleigh wave, since the pillar vibrates in the out of plane direction, and used the corresponding SAW velocity $c_\text{Ray}$ given in the main text. The determined $\theta$ is shown in Fig.~\ref{FEMSinglePillar}c. It first rises at small distances $r$ and then reaches a constant value of around $50^\circ$ after approximately two SAW wavelengths. A constant $\theta$ means, that the SAW propagates with the phase velocity used to calculate $\theta$. This verifies our assumption that the pillar resonator emits Rayleigh waves. The fact, that $\theta$ is not constant from the very beginning, but rises at small distances $r$, supports the argumentation, that a SAW, emitted by a vibrating pillar in the first bulk mode, undergoes an additional phase chance during its formation.  

\subsection*{3. FEM simulations of the second bending mode}
To test the SAW coupling model further, we also simulated two pillar resonators vibrating in the second bending mode. For the simulations, we adapted the simulations of the first bulk mode and made two changes. The first change was the replacement of the symmetric boundary condition by the following antisymmetric boundary condition
 \begin{equation}
\mathbf{u} \cdot \mathbf{t_1} = 0 \And \mathbf{u} \cdot \mathbf{t_2} = 0  \; ,
\label{BCantisym}
\end{equation}
 where $\mathbf{t_1}$ and $\mathbf{t_2}$ are two perpendicular vectors in the considered sectional plane. In our case, this is the yz-plane. A pillar resonator made out of an isotropic material shows only a displacement in x-direction at any point in the yz-plane when vibrating in the second bending mode in x-direction in the limit of small vibrational amplitudes \cite{Schmid2016,Weaver1990}. 
 
\begin{figure}[t]
  \begin{center}    
    \includegraphics{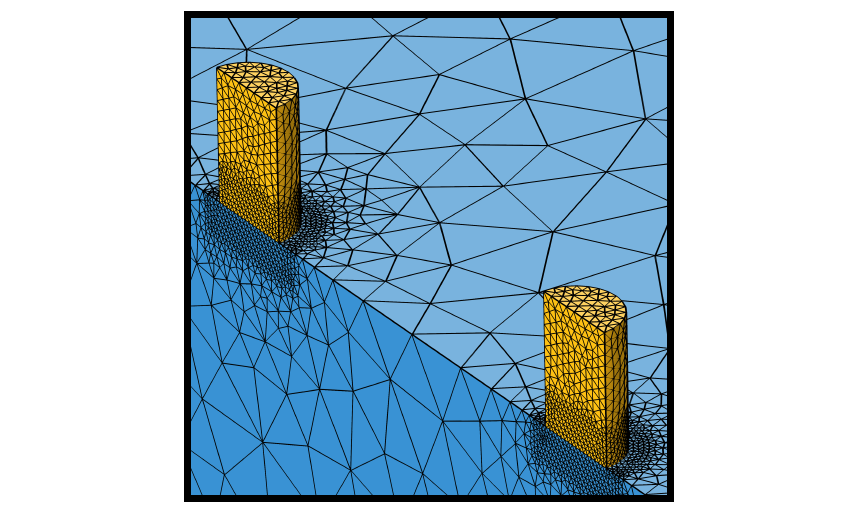} 
    \caption{Detail of the geometry and mesh of the FEM simulations of two pillar resonators vibrating in the second bending mode. The distance between the pillars is ${d = \lambda_\text{SAW}/2}$ with $\lambda_\text{SAW} = c_\text{y}/f_0$.}
    \label{MeshBending}
  \end{center}    
\end{figure}

The second change was the reduction of the maximum element size of the unstructured tetrahedral mesh around the pillars' base by a factor of two, as shown in Fig.~\ref{MeshBending}. We refined the mesh to reduce the dependency of the simulated eigenfrequencies on the mesh.

\begin{figure}[t]
  \begin{center}    
    \includegraphics{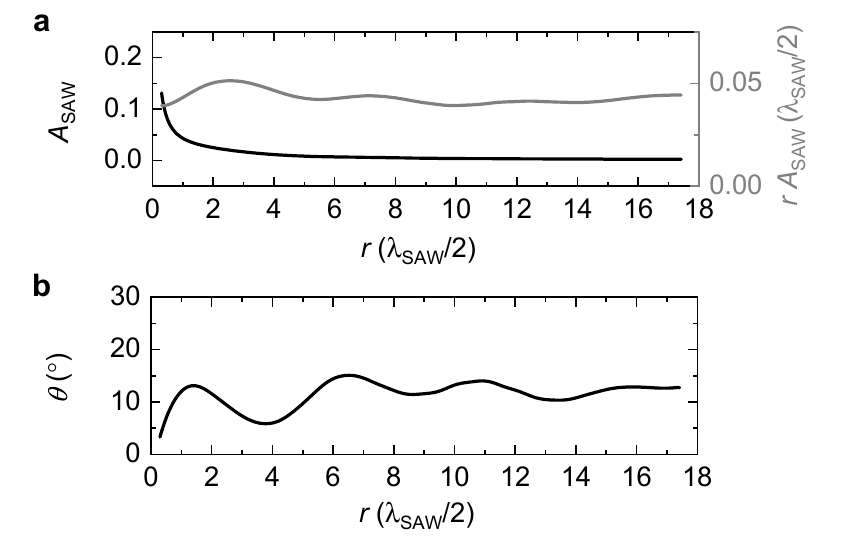} 
    \caption{FEM simulation of a single pillar resonator vibrating in the second bending mode. \textbf{a} Normalized amplitude $A_\text{SAW}$ \big($A_\text{SAW}(0) = 1$\big) of a SAW emitted by a single pillar resonator after propagating a distance $r$ along the y-direction. \textbf{b} Additional phase difference $\theta$ between the phase of the pillar resonator and the phase of the pillar's emitted SAW at distances $r$ to the pillar.}
    \label{FEMSinglePillarBending}
  \end{center}    
\end{figure}

\begin{figure}[t]
  \begin{center}    
    \includegraphics{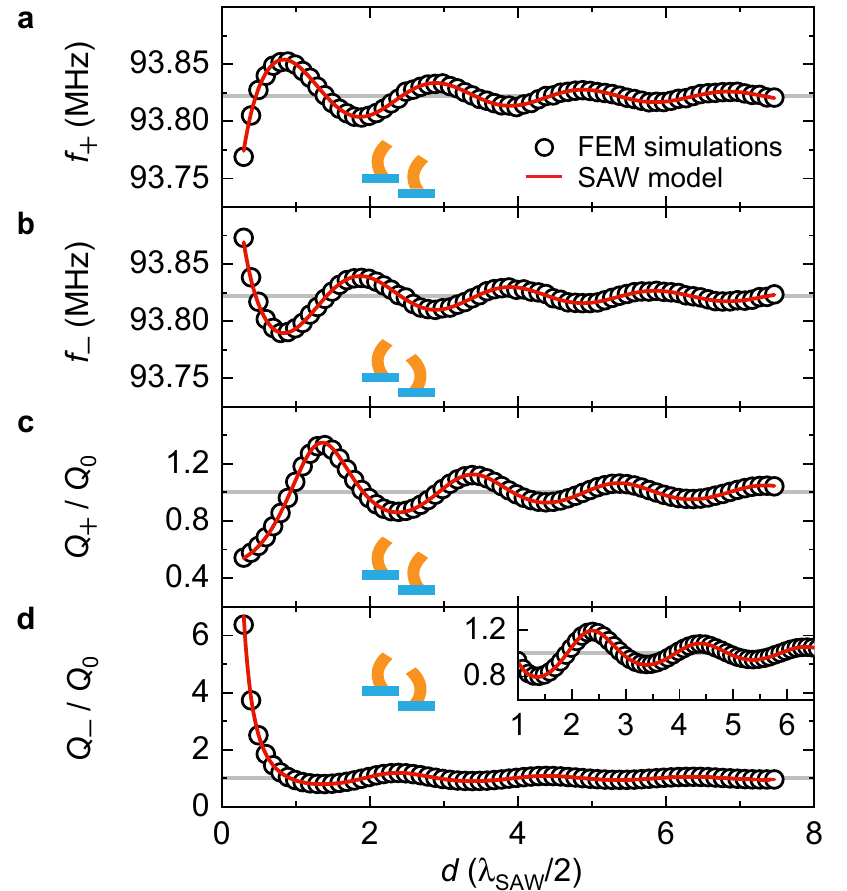}
    \caption{FEM simulations of a pair of pillar resonators vibrating in the second bending mode. Eigenfrequency and quality factor of the symmetric and antisymmetric modes as a function of the distance $d$ between the pillars. The grey lines mark the properties $f_0 = 93.82~\text{MHz}$ and $Q_0 = 534$ of a single pillar resonator. The red lines are plots of $\omega_\pm(d)$ and $Q_\pm(d)$, given in the main text, for $g=0.014$ and exploiting $\omega_\pm \approx \omega_0$.}
    \label{FEM2PillarsBending}
  \end{center}    
\end{figure}

\begin{figure}[t]
  \begin{center}    
    \includegraphics{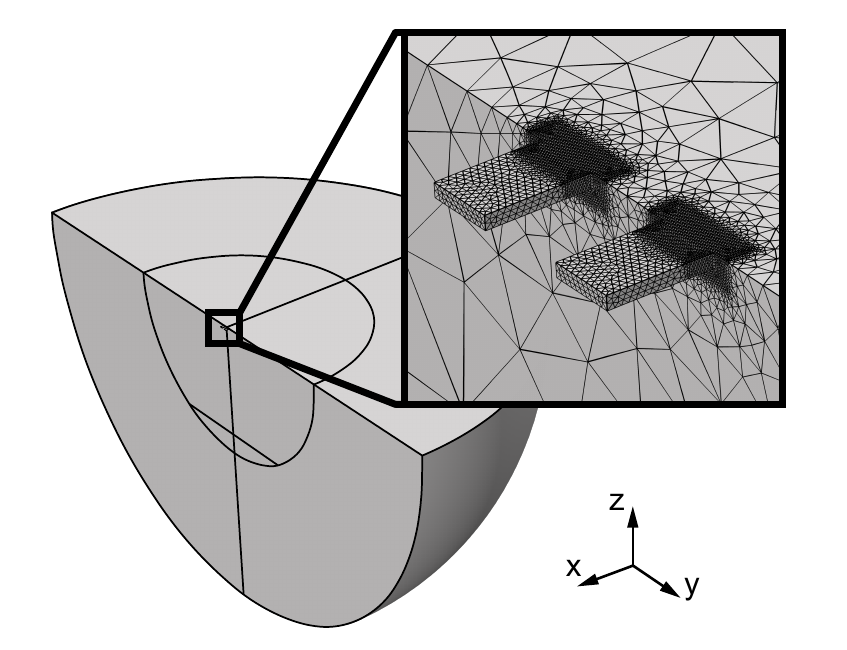} 
    \caption{Geometry and mesh of the FEM simulations of two cantilevers for a separation distance of $d = 0.1 \,\lambda_\text{SAW}/2$ with $\lambda_\text{SAW} = c_\text{Ray}/f_0$.}
    \label{GeometryCantilever}
  \end{center}    
\end{figure}

The results of the single pillar FEM simulation are shown in Fig.~\ref{FEMSinglePillarBending}. The results were obtained in the same way as described for the simulation of the first bulk mode with three exceptions. We determined $A_\text{SAW}$ and $\Delta \phi$ based on the displacement in x-direction $u_\text{x}$ and not z-direction $u_\text{z}$, since the pillars vibrate in a bending mode in x-direction. For the calculation of $\theta$, we added $180^\circ$ to the resonator phase $\phi_\text{R}$ due to the phase difference between the pillar's base and its top, and we didn't use the Rayleigh wave velocity $c_\text{Ray}$ to determine $\theta$, but estimated the phase velocity by a linear fit of $\Delta \phi$ ($c_\text{y}=\SI{4080}{\meter / \second}$). The reason for the latter was that pillar resonators vibrating in a bending mode in x-direction do not emit Rayleigh waves in the y-direction, since they don't show any displacement in the z-direction in the yz-plane \cite{Schmid2016,Weaver1990}. This argumentation is supported by Fig.~\ref{FEMSinglePillarBending}a. It can be seen, that the amplitude along the surface in y-direction is proportional to the propagation distance $r$, equivalent to the case of a point source surrounded by an isotropic medium. This means, that the corresponding wave emitted by the pillar does not only propagate on the surface, but also into the substrate. Consequently, it is not a pure SAW \cite{Morgan2007}. This might be the reason for the oscillations in $\theta$, which is shown in Fig.~\ref{FEMSinglePillarBending}b. Waves are propagating away from the surface, but with a different group velocity than the waves propagating along the surface.

\begin{figure}[t]
  \begin{center}    
    \includegraphics{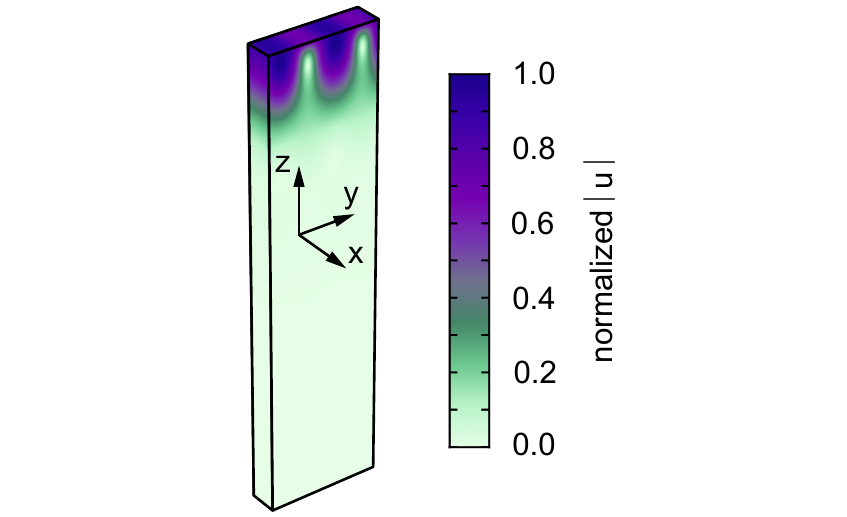} 
    \caption{FEM simulation to determine the Rayleigh wave velocity in crystalline silicon along the [100]-direction. A propagating Rayleigh wave is shown. The colors indicate the normalized total displacement.}
    \label{GeometryPhaseVelocity}
  \end{center}    
\end{figure}

\begin{figure}[t]
  \begin{center}    
    \includegraphics{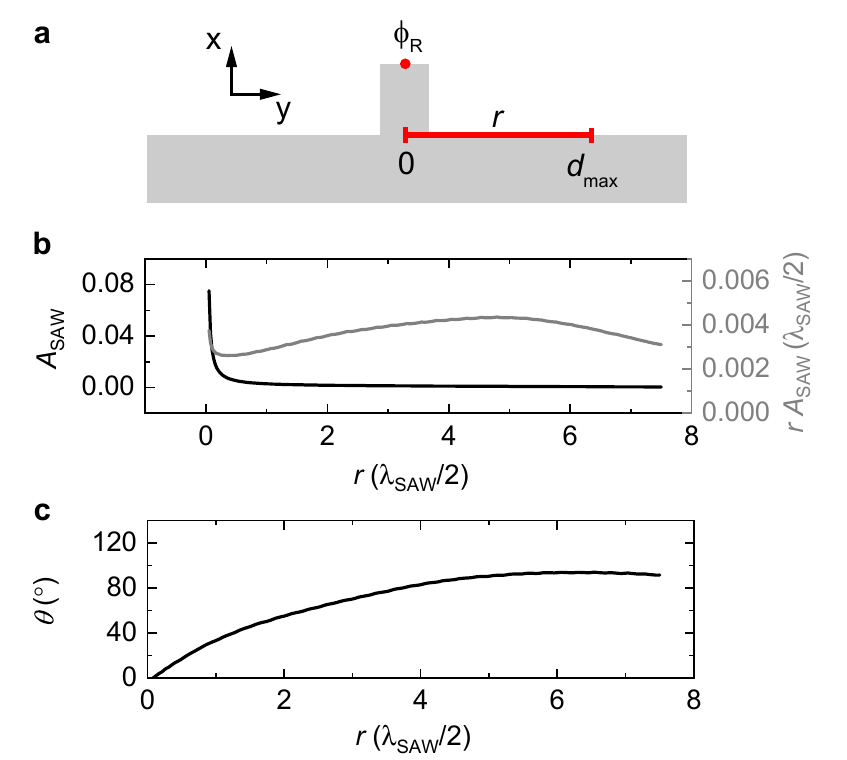} 
    \caption{FEM simulation of a single cantilever. \textbf{a} Schematic of a single cantilever mounted on a substrate illustrating along which line the normalized amplitude $A_\text{SAW}$ and the additional phase difference $\theta$ of the cantilever's emitted SAW were determined. The parameter $d_\text{max}$ is the maximum distance between two cantilevers in the FEM simulations of a pair of cantilevers. The red dot marks the spot at which the phase of the resonator $\phi_\text{R}$ was determined. \textbf{b} Normalized amplitude $A_\text{SAW}$ \big($A_\text{SAW}(0) = 1$\big) of a SAW emitted by a single cantilever after propagating a distance $r$ along the y-direction. \textbf{c} Additional phase difference $\theta$ between the phase of the cantilever and the phase of the cantilever's emitted SAW at distances $r$ to the cantilever.}
    \label{FEMSingleCantilever}
  \end{center}    
\end{figure}

The results of the FEM simulations of two pillar resonators are shown in Fig.~\ref{FEM2PillarsBending} and show excellent agreement with the proposed SAW coupling model. We used the same fitting procedure as described for the simulations of the first bulk mode (see main text). 

\subsection*{4. FEM simulations of two cantilevers}

To test the SAW coupling model further, we simulated two MHz frequency cantilevers vibrating in the first bending mode. The cantilevers were etched into a (100) silicon substrate with the crystallographic [110]-direction aligned with the geometric x-axis. A schematic of the geometry is depicted in Fig.~\ref{GeometryCantilever}. The cantilevers had a length of \SI{1.5}{\micro\meter}, a width of \SI{1}{\micro\meter} and a height of \SI{0.2}{\micro\meter}, similar to the dimensions of published high frequency cantilevers\cite{Li2007}. The substrate was modeled by a quarter sphere and partitioned into an outer and an inner part. The outer part was defined as a perfectly matched layer (PML) mimicking an infinitive large substrate.

An unstructured tetrahedral mesh was used for the cantilevers with a maximum element size of an half of the cantilevers thickness. Around the anchors, we refined the mesh by a factor of two, as can be seen in Fig.~\ref{GeometryCantilever}, to reduce the dependency of the simulated eigenfrequencies on the mesh. A swept mesh was utilized for the PML and an unstructured tetrahedral mesh for the inner part of the substrate. The latter had a maximum element size of an eight of the SAW wavelength $\lambda_\text{SAW}$ up to a distance of $\lambda_\text{SAW}$ to the surface. For elements deeper in the substrate, the maximum element size was reduced by a factor of two. For the calculation of $\lambda_\text{SAW}$, we used the Rayleigh wave velocity in silicon along the [110]-direction or geometric x-direction, which is also the Rayleigh wave velocity in the geometric y-direction due to the cubic symmetry of silicon. By simulating the eigenmodes of a silicon block with the same crystallographic orientation like the considered silicon substrate, we got $c_\text{Ray} = \SI{5092}{\meter / \second}$. The geometry of the block and the simulated Rayleigh wave are shown in Fig.~\ref{GeometryPhaseVelocity}. We applied periodic boundary conditions to the xz- and yz-faces and a fixed constraint to the bottom xy-face. The Rayleigh wave velocity is then given by the eigenfrequency of the corresponding mode times the length of the block in y-direction.

In comparison to the FEM simulations of the pillar resonators, we didn't use a symmetric or antisymmetric boundary condition for the FEM simulations of the cantilevers, but implemented a symmetry in the mesh. We first build the mesh on one half of the structure and then copied the mesh to the other half. Without that, the simulated modes were not fully symmetric or antisymmetric, but the phase difference between the two cantilevers significantly differed from $0^\circ$ or $180^\circ$ for some separation distances between the cantilevers.
\begin{figure}[t]
  \begin{center}    
    \includegraphics{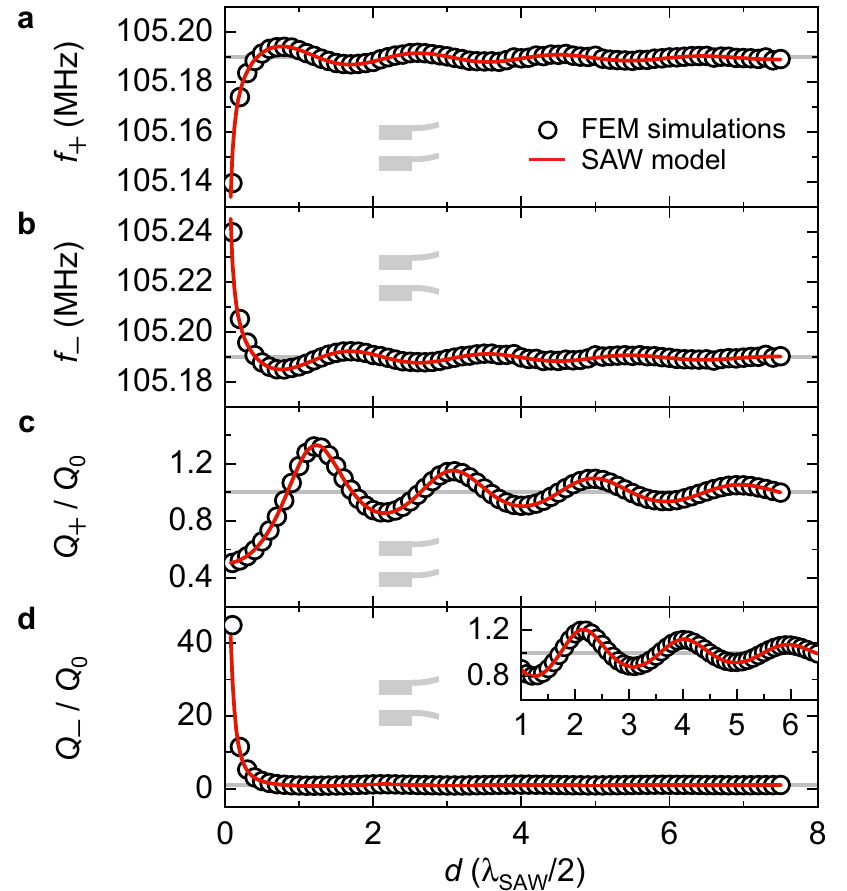}
    \caption{FEM simulations of two cantilevers. Eigenfrequency and quality factor of the symmetric and antisymmetric modes as a function of the distance $d$ between the cantilevers. The grey lines mark the properties $f_0 = 105.19~\text{MHz}$ and $Q_0 = 3991$ of a single cantilever. The red lines are plots of $\omega_\pm(d)$ and $Q_\pm(d)$, given in the main text, for $g=0.026$ and exploiting $\omega_\pm \approx \omega_0$.}
    \label{FEM2Cantilevers}
  \end{center}    
\end{figure}
We used the same shape functions as in the FEM simulations of the pillar resonators and also an eigenfrequency study to calculate the eigenfrequencies and quality factors of the symmetric and antisymmetric normal modes for different distances $d$.

The results of the FEM simulation of a single cantilever are shown in Fig.~\ref{FEMSingleCantilever}. The results were obtained in the same way as described for the simulation of the single pillar resonator vibrating in the first bulk mode. A difference between the pillars and the cantilevers is, that in case of the cantilevers the emitted SAW, which is interacting with the other resonator, is not propagating along a closed surface, but on a edge. Consequently, this SAW has a more complicated propagation behaviour than a pure SAW, which is also indicated by the normalized amplitude $A_\text{SAW}$ shown in Fig.~\ref{FEMSingleCantilever}b. For the calculation of $\theta$, we guessed the velocity of the SAW propagating along the edge and used the simulated Rayleigh wave velocity $c_\text{Ray} = \SI{5092}{\meter / \second}$. Considering Fig.~\ref{FEMSingleCantilever}c, it seems, that $c_\text{Ray}$ is not too far away from the actual wave velocity in the given range, since $\theta$ does not constantly raise or drop.

The results of the FEM simulations of two cantilevers are shown in Fig.~\ref{FEM2Cantilevers} and show excellent agreement with the proposed SAW coupling model. We used the same fitting procedure as described for the simulations of the first bulk mode (see main text). The smallest separation distance shown in Fig.~\ref{FEM2Cantilevers} is $d = 0.1 \,\lambda_\text{SAW}/2$. In comparison, the smallest separation distance in the pillar simulations is $d = 0.3 \,\lambda_\text{SAW}/2$. The reason for the difference is, that the pillars are wider than the cantilevers compared to the SAW wavelength.

\clearpage

\end{document}